\documentclass[superscriptaddress,twocolumn,showpacs,preprintnumbers]{revtex4}
\usepackage{eurosym}
\usepackage[tbtags]{amsmath}
\usepackage{amsfonts}
\usepackage{mathrsfs}
\usepackage{amssymb}
\usepackage{graphicx}
\usepackage{epsfig,graphicx,times}
\usepackage{subfigure}
\usepackage{color}
\usepackage{amssymb}
\usepackage{amsmath}
\usepackage{graphicx}
\usepackage{dcolumn}
\usepackage{bm}
\usepackage{float}
\usepackage[abs]{overpic}
\usepackage{epstopdf}

\input{tcilatex}
\begin{document}

\title{Phase-dependent optical response properties in an optomechanical system by coherently driving the mechanical resonator}
\author{W. Z. Jia}
\email{wenzjia@home.swjtu.edu.cn}
\affiliation{Quantum Optoelectronics Laboratory, School of Physical Science and Technology,
Southwest Jiaotong University, Chengdu 610031, China}
\author{L. F. Wei}
\affiliation{Quantum Optoelectronics Laboratory, School of Physical Science and Technology,
Southwest Jiaotong University, Chengdu 610031, China}
\affiliation{State Key Laboratory of Optoelectronic Materials and Technologies,
School of Physics Science and Engineering, Sun Yet-sen University, Guangzhou 510275, China}
\author{Yong Li}
\affiliation{Beijing Computational Science Research Center, Beijing 100084, China}
\affiliation{Synergetic Innovation Center of Quantum Information and Quantum Physics, University of Science and Technology of China, Hefei, Anhui 230026, China}
\author{Yu-xi Liu}
\email{yuxiliu@mail.tsinghua.edu.cn}
\affiliation{Institute of Microelectronics, Tsinghua University, Beijing 100084, China}
\affiliation{Tsinghua National Laboratory for Information Science and Technology (TNList), Beijing 100084, China}
\date{\today }

\begin{abstract}
We explore theoretically the optical response properties in an optomechanical system under electromagnetically induced transparency condition but with the mechanical resonator being driven by an additional coherent field. In this configuration, more complex quantum coherent and interference phenomena occur. In particular, we find that the probe transmission spectra depend on the total phase of the applied fields. Our study also provides an efficient way to control propagation of a probe field from perfect absorption to remarkable
amplification.
\end{abstract}

\maketitle

\bigskip




\bigskip

\section{\label{p1}Introduction}

Optomechanical systems couple photons and phonons via radiation pressure.
Significant research interest in this frontier of optomechanics is motivated
by its potential applications in ultrasensitive measurements, quantum
information processing, and implementation of novel quantum phenomena at
macroscopic scales \cite{review1,review2,review3,review4}.
Recently, a phenomenon resembling electromagnetically induced transparency
(EIT) \cite{EIT1,EIT2,EIT3} in atomic physics, called optomechanically induced transparency (OMIT),
is studied theoretically \cite{OMIT1a,OMIT1b,OMIT1c} and observed experimentally \cite{OMIT2,OMIT3,OMIT4,OMIT5}. OMIT can be used for slowing and switching probe signals \cite{OMslowlight1} and may be further used for
 on-chip storage of light pulses via microfabricated
optomechanical arrays \cite{OMslowlight2}. OMIT in the nonlinear quantum regime has also been investigated \cite{nolinearOMIT1,nolinearOMIT2,nolinearOMIT3,nolinearOMIT4}.  On the other hand, 
optomechanically induced absorption (OMIA) phenomenon, which is an analog of
electromagnetically induced absorption (EIA) investigated in atomic gas \cite{EIA1,EIA2} and superconducting 
artificial atoms \cite{EIA3}, can also be
realized in optomechanical setup \cite{OMIT4,OMIA1,OMIA2}. And OMIA is a phenomenon closely related to optomechanically induced amplification \cite{OMIT4,OMIT5,OMAmplification1,OMAmplification2,OMAmplification3}.

\begin{figure}[t]
\includegraphics[width=0.5\textwidth]{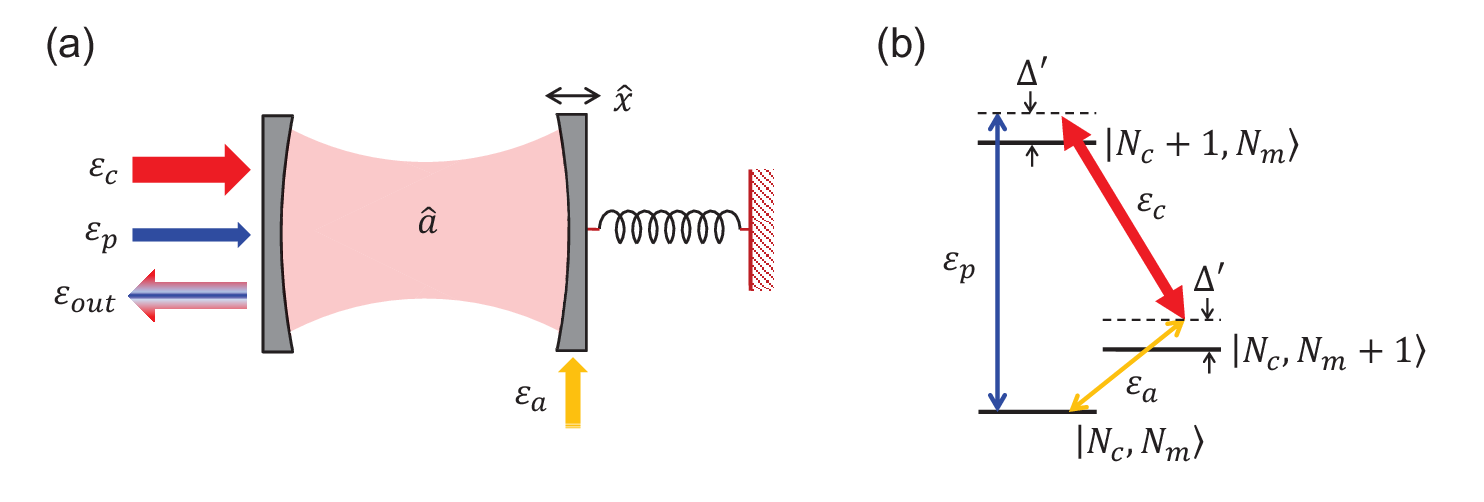}
\caption{(color online). (a) Standard optomechanical setup driven by a
control laser and a probe laser, with an auxiliary driving field applied to the mechanical resonator. (b) A block of three energy levels in the system. 
$N_{c}$ ($N_{m}$) denotes the number of photons (phonons). The control laser
with amplitude $\protect\varepsilon _{\mathrm{c}}$ resonantly couples the transition 
$\left\vert N_{c},N_{m}+1\right\rangle \leftrightarrow\left\vert N_{c}+1,N_{m}\right\rangle $, 
the probe laser with amplitude $\protect
\varepsilon _{\mathrm{p}}$ and detuning $\Delta^{\prime}$ couples the transition 
$\left\vert N_{c},N_{m}\right\rangle \leftrightarrow\left\vert
N_{c}+1,N_{m}\right\rangle $, the auxiliary driving field with amplitude $\protect
\varepsilon _{\mathrm{a}}$ and detuning $\Delta^{\prime}$ couples the transition 
$\left\vert N_{c},N_{m}\right\rangle \leftrightarrow\left\vert
N_{c},N_{m}+1\right\rangle $.}
\label{OM system}
\end{figure}

To abtain optomechanical analogs of atomic coherence related phenomena such
as EIT and EIA, the key point is that a mechanical coherence (similar to
atomic coherence) must be induced. Specifically, in standard OMIT \cite{OMIT1a,OMIT1b,OMIT1c,OMIT2,OMIT3,OMIT4,OMIT5} and OMIA \cite{OMIT4,OMIA1},
the coherent oscillation of the mechanical resonator results from a time varying
radiation pressure force induced by the beat of the probe and the control
laser. The oscillating mechanical resonator together with (red-/blue-detuned) control field can further induce sidebands on the cavity field.
The generated field with probe frequency can interfer with the original
probe field, leading to OMIT/OMIA absorption spectra.
On the other hand, in three-level atomic physics, the atomic coherences
can be produced by the direct drive at the microwave
frequency \cite{AtomloopScully1,AtomloopScully2,AtomloopScully3} or by the spontaneously generated
coherence \cite{SGC}.  Usually, these additional drives can generate
closed transition loop. These so-called phaseonium systems \cite{phaseonium1,phaseonium2}
can lead to many remarkable phase dependent effects such as correlated lasing \cite{CorrelatedLasing1,CorrelatedLasing2} and 
inversionless gain \cite{AtomloopScully1,AtomloopScully2,AtomloopKnight}.
Similarly, in optomechanical system, one can expect that this
type of mechanical coherence can also be generated by directly 
driving the mechanical resonator, and further used to coherent control the propagation of
probe fields. Thus, in this paper, we study the influence of directly produced mechanical coherence on optical response properties of an
optomechanical system. 

In our study,
besides a red-detuned control field and a nearly resonant probe field applied to pump the optical cavity, an additional weak driving field is used to
directly excite the mechanical resonator to produce mechanical coherence.
In contrast to the strong magnetic driving used for coherent connection between two electric-dipole-forbidden
atomic energy levels\cite{AtomloopScully1,AtomloopScully2,AtomloopScully3},
the weak electric driving is enough for our study here, because there is no
selection rule in our system.
 In this case, the optomechnical cavity can be resonantly exited by directly absorbing
a probe photon, or through phonon-photon process. For the interference effects of these two
possible transition paths, the optical response properties
for the probe field become phase-sensitive, and more complex quantum
interference and quantum coherence related phenomena will appear. Specifically, gain without inversion (GWI) like, OMIA and
EIT-type spectra can be abtained, depending on the amplitude and phase of
the control field. In addition, by adjusting the control field and the additional
driving field applied on the mechanical resonator, the probe field can
be efficiently manipulated from perfect absorption to remarkable
amplification.

The paper is organized as follows. In Sec.~\ref{p2}, we introduce theoretical model
for describing the driven optomechanical system. Then, in Sec.~\ref{p3}, we
study the phase-dependent optical response for the probe field in detail,
including GWI-like spectra in Sec.~\ref{A}, OMIA and EIT-like
spectra in Sec.~\ref{B}, amplification and perfect absorption in Sec.~\ref{C}, and
numerical simulation in Sec.~\ref{D}. Finally, further discussions and
conclusions are given in Sec.~\ref{p4}.

\section{\label{p2}The model}

We consider a standard optomechanical system schematically illustrated in
Fig.~\ref{OM system}(a). The cavity is driven by a strong control laser and
a weak probe one, where $\omega _{\mathrm{c}}$ ($\omega _{\mathrm{p}}$)
and $\varepsilon _{\mathrm{c}}$ ($\varepsilon _{\mathrm{p}}$) are the control
(probe) laser frequency and amplitude, respectively. Meanwhile, a weak
coherent driving field with frequency $\omega _{\mathrm{a}}$ and amplitude $
\varepsilon _{\mathrm{a}}$ is applied to excite the mechanical resonator.
Experimentally, a micro/nano scale mechanical
resonator can be driven by microwave electrical signals \cite{OMdr1,OMdr2,OMdr3}. For
example, in a recent experiment \cite{OMdr1}, the mechanical element is a thin film of piezoelectric materials AlN, which is sandwiched between two aluminium metal electrodes, enabling strong electromechanical coupling through the piezoelectric effect. 
 We also assume that the frequencies of the three coherent driving
fields satisfy the condition $\omega _{\mathrm{p}}-\omega _{\mathrm{c}}=\omega _{\mathrm{a}
}$. Fig.~\ref{OM system}(b) shows a block of three energy levels in the
system. Clearly, the three couplings create a set of $\Delta $-type
transitions analogous to those in 
microwave-driving natural atoms \cite{AtomloopScully1,AtomloopScully2,AtomloopScully3,AtomloopKosachiov}, superconducting artificial atoms \cite{DeltaAA1,DeltaAA2,DeltaAA3,DeltaAA4}, or chiral
molecules \cite{DeltaMolecule1,DeltaMolecule2}. Thus one can expect that, similar to these quantum systems with closed-loop transition structure, the optical properties of the optomechanical system considered here will be sensitive to the relative phases of three applied fields.

In a frame rotating at the frequency of the coupling field $\omega _{\mathrm{
c}}$, The Hanmiltonian of the system is of the form
\begin{equation}
\hat{H}=\hbar \Delta _{0}\hat{c}^{\dagger }\hat{c}+\hbar \omega _{\mathrm{m}}
\hat{b}^{\dagger }\hat{b}-\hbar g_{0}\hat{c}^{\dagger }\hat{c}\left( \hat{b}
^{\dagger }+\hat{b}\right) +\hat{H}_{\mathrm{dr}},  
\label{Hamiltonian}
\end{equation}
where $\hat{c}$ ($\hat{b}$) is the photon (phonon) annihilation operator, $
\omega _{\mathrm{m}}$ is the mechanical resonance frequency, $\Delta
_{0}=\omega _{0}-\omega _{\mathrm{c}}$ is the detuning of the control laser from the
bare cavity frequency $\omega _{0}$, $g_{0}$ is the single-photon coupling
strength of the radiation pressure between the cavity field and the mechanical resonator, and $
\hat{H}_{\mathrm{dr}}$ describes the interaction between the optomechanical system and the three driving fields:
\begin{equation}
\hat{H}_{\mathrm{dr}}=i\hbar \left( \varepsilon _{\mathrm{c}}+\varepsilon _{%
\mathrm{p}}e^{-i\omega _{\mathrm{a}}t}\right) \hat{c}^{\dagger }+i\hbar
\varepsilon _{\mathrm{a}}e^{-i\omega _{\mathrm{a}}t}\hat{b}^{\dagger }+%
\mathrm{H.c.}  
\label{Driving}
\end{equation}
The nonlinear quantum Langevin equations for the operators of the optical
and mechanical modes are given by
\begin{align}
\dot{\hat{c}}& =-\left( i\Delta _{0}+\frac{\kappa }{2}\right) \hat{c}+ig_{0}
\hat{c}\left( \hat{b}^{\dagger }+\hat{b}\right) +\varepsilon _{\mathrm{c}
}+\varepsilon _{\mathrm{p}}e^{-i\omega _{\mathrm{a}}t}+\hat{f},
\label{Langevin1}
 \\
\dot{\hat{b}}& =-\left( i\omega _{\mathrm{m}}+\frac{\gamma _{\mathrm{m}}}{2}
\right) \hat{b}+ig_{0}\hat{c}^{\dagger }\hat{c}+\varepsilon _{\mathrm{a}
}e^{-i\omega _{\mathrm{a}}t}+\hat{\xi}. 
\label{Langevin2}
\end{align}
$\kappa $ and $\gamma _{\mathrm{m}}$ are the decay rates of cavity and
mechanical resonator, respectively. $\hat{f}$ and $\hat{\xi}$ are the quantum and
thermal noise operators, respectively. We assume that they satisfy the condition $\langle \hat{f}\rangle =\langle \hat{
\xi}\rangle =0$.

It is not easy to obtain the solutions of the nonlinear equations \eqref{Langevin1} and \eqref{Langevin2}.
However, we are only interested in the linear response of the driven optomechanical system
to weak probe field. Thus, in the case of $|\varepsilon_\mathrm{p}|,\,|\varepsilon_\mathrm{a}| 
\ll |\varepsilon_\mathrm{c}|$, we can linearize the dynamical equations of the driven optomechanical
system by assuming $\hat{c}=c_{\mathrm{s}}+\delta \hat{c}$ and $\hat{b}=b_{\mathrm{s}}+\delta \hat{b}$. Here $c_{\mathrm{s}}$ and $b_{\mathrm{s}}$
are steady-state values of the system when only strong driving field is applied. They can be gotten from
Eqs.~\eqref{Langevin1} and \eqref{Langevin2} by assuming $\varepsilon _{\mathrm{p}},\varepsilon _{\mathrm{a}
}\rightarrow 0$ and all time derivatives vanish: 
\begin{equation}
c_{\mathrm{s}}=\frac{\varepsilon _{\mathrm{c}}}{i\Delta +\frac{\kappa }{2}}
,b_{\mathrm{s}}=\frac{ig_{0}\left\vert c_{\mathrm{s}}\right\vert ^{2}}{
i\omega _{\mathrm{m}}+\frac{\gamma _{\mathrm{m}}}{2}},
\end{equation}
where $\Delta =\Delta _{0}-g_{0}\left( b_{\mathrm{s}}+b_{\mathrm{s}}^{\ast
}\right) $ denotes the effective detuning between the cavity field and the
control field, including the frequency shift caused by the mechanical
motion. After plugging the ansatz $\hat{c}=c_{\mathrm{s}}+\delta \hat{c},
\hat{b}=b_{\mathrm{s}}+\delta \hat{b}$ into Eqs.~\eqref{Langevin1} and
\eqref{Langevin2}, and dropping the small nonlinear terms, we can get the
linearized quantum Langevin equations for the operators $\delta \hat{c}$ and $\delta \hat{b}$: 
\begin{align}
\dot{\delta \hat{c}}& =-\left( i\Delta +\frac{\kappa }{2}\right) \delta \hat{
c}+iG\left( \delta \hat{b}^{\dagger }+\delta \hat{b}\right) +\varepsilon _{
\mathrm{p}}e^{-i\omega _{\mathrm{a}}t}+\hat{f}, 
\\
\dot{\delta \hat{b}}& =-\left( i\omega _{\mathrm{m}}+\frac{\gamma _{\mathrm{m
}}}{2}\right) \delta \hat{b}+i\left( G\delta \hat{c}^{\dagger }+G^{\ast
}\delta \hat{c}\right) +\varepsilon _{\mathrm{a}}e^{-i\omega _{\mathrm{a}}t}+
\hat{\xi},
\end{align}
where $G=g_{0}c_{\mathrm{s}}$ is the total (linearized optomechanical)
coupling strength. 

Now we move into another interaction picture by
introducing $\delta \hat{c}\rightarrow \delta \hat{c}e^{-i\omega _{\mathrm{a}
}t}$, $\delta \hat{b}\rightarrow \delta \hat{b}e^{-i\omega _{\mathrm{a}}t}$, 
$\hat{f}\rightarrow \hat{f}e^{-i\omega _{\mathrm{a}}t},\hat{\xi}\rightarrow 
\hat{\xi}e^{-i\omega _{\mathrm{a}}t}$. In addition, we assume the cavity is
driven by a control field at the mechanical red sideband with $\Delta =\omega _{
\mathrm{m}}$, the system is operating in the resolved sideband regime $
\omega _{\mathrm{m}}/ \kappa\gg1$, the mechanical resonator has a high mechanical quality
factor $\omega _{\mathrm{m}}/\gamma _{\mathrm{m}}\gg1$, and the mechanical
frequency $\omega _{\mathrm{m}}$ is much larger than $\left\vert
G\right\vert $ and $\left\vert \omega _{\mathrm{a}}-\omega _{\mathrm{m}
}\right\vert $. In this parameter regime, analogous to the rotating
wave approximation presented in the context of atomic EIT, one can ignore
the fast oscillating terms $e^{2i\omega _{\mathrm{a}}t}$ and get the
following equations: 
\begin{align}
\dot{\delta \hat{c}}& =\left( i\Delta ^{\prime }-\frac{\kappa }{2}\right)
\delta \hat{c}+iG\delta \hat{b}+\varepsilon _{\mathrm{p}}+\hat{f},
\label{linearLangevinRWA1} 
\\
\dot{\delta \hat{b}}& =\left( i\Delta ^{\prime }-\frac{\gamma _{\mathrm{m}}}{
2}\right) \delta \hat{b}+iG^{\ast }\delta \hat{c}+\varepsilon _{\mathrm{a}}+
\hat{\xi},  
\label{linearLangevinRWA2}
\end{align}
with $\Delta ^{\prime }=\omega _{\mathrm{a}}-\omega _{\mathrm{m}}=\omega _{
\mathrm{p}}-\omega _{\mathrm{c}}-\omega _{\mathrm{m}}$.
Then we take the expectation values of the operators in Eqs.~\eqref{linearLangevinRWA1} and
\eqref{linearLangevinRWA2}. Note that the mean values of the quantum and thermal noise
terms are zero (i.e., $\langle \hat{f}\rangle =\langle \hat{
\xi}\rangle =0$). Under steady-state condition $\langle\dot{\delta \hat{c}}\rangle=\langle\dot{\delta \hat{b}}\rangle=0$, one has
\begin{align}
0& =\left( i\Delta ^{\prime }-\frac{\kappa }{2}\right)
\langle\delta\hat{c}\rangle+iG\langle\delta \hat{b}\rangle+\varepsilon _{\mathrm{p}},
\label{exp1} 
\\
0& =\left( i\Delta ^{\prime }-\frac{\gamma _{\mathrm{m}}}{
2}\right) \langle\delta \hat{b}\rangle+iG^{\ast }\langle\delta\hat{c}\rangle+\varepsilon _{\mathrm{a}}.  
\label{exp2}
\end{align}
\begin{figure*}[t]
\includegraphics[width=1
\textwidth]{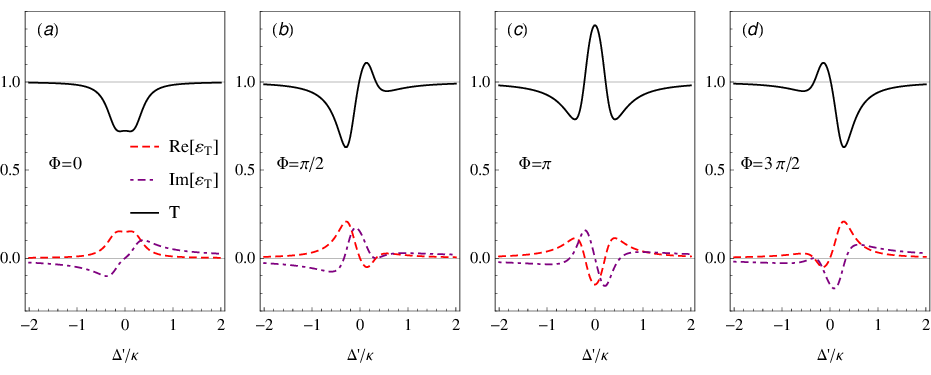}
\caption{(color online). Phase-dependent absorption (dashed line),
dispersion (dash-dotted line), and power transmission coefficient (solid
line) versus $\Delta ^{\prime }$ for different phase factor: (a) $\Phi =0$;
(b) $\Phi =\protect\pi /2$; (c) $\Phi =\protect\pi $; (d) $\Phi =3\protect
\pi /2$. Other parameters are $\left\vert G\right\vert=\protect\kappa /3$, $\protect\omega _{
\mathrm{m}}=10\protect\kappa $, $\protect\gamma _{\mathrm{m}}=\protect\kappa 
/1000$, $\protect\eta =0.05$, $y=1$.}
\label{EITA}
\end{figure*}
Thus, the expectation value of the operator $
\delta \hat{c}$ corresponding to intra-cavity field oscillating at the probe
frequency reads 
\begin{eqnarray}
\left\langle \delta \hat{c}\right\rangle &=&e^{i\phi _{\mathrm{p}}}\left[ 
\frac{\left( \frac{\gamma _{\mathrm{m}}}{2}-i\Delta ^{\prime }\right)
\left\vert \varepsilon _{\mathrm{p}}\right\vert }{\left( \frac{\kappa }{2}%
-i\Delta ^{\prime }\right) \left( \frac{\gamma _{\mathrm{m}}}{2}-i\Delta
^{\prime }\right) +\left\vert G\right\vert ^{2}}\right.  \notag \\
&&\left. +\frac{\left\vert \varepsilon _{\mathrm{a}}\right\vert \left\vert
G\right\vert e^{i\Phi }}{\left( \frac{\kappa }{2}-i\Delta ^{\prime }\right)
\left( \frac{\gamma _{\mathrm{m}}}{2}-i\Delta ^{\prime }\right) +\left\vert
G\right\vert ^{2}}\right].  
\label{expectationC}
\end{eqnarray}
Here the total phase $\Phi$ is defined as $\arctan \left( \frac{\kappa 
}{2\omega _{\mathrm{m}}}\right) +\phi _{\mathrm{c}}+\phi _{\mathrm{a}}-\phi
_{\mathrm{p}}$, $\phi _{\mathrm{i}}$ is the phase of amplitude $
\varepsilon _{\mathrm{i}}$ ($\mathrm{i=c,a,p}$). In the resolved sideband
limit, $\Phi \simeq \phi _{\mathrm{c}}+\phi _{\mathrm{a}}-\phi _{\mathrm{p}}$. In Eq.~\eqref{expectationC}, the first term is the contribution
from usual OMIT effect \cite{OMIT1a,OMIT2}, and the second term is the contribution from the phonon-photon
parametric process involving the driving on the mechanical resonator. The intra-cavity field with probe frequency
 is determined by the interference of these two terms and is
strongly dependent on the relative phase of the applied driving fields. Thus we can
control the transmission of the probe field by adjusting
the total phase $\Phi$.

The output field of the cavity can be derived by the input-output relation \cite{inputoutput}
\begin{equation}
\left\langle \hat{c}_{\mathrm{out}}\right\rangle +\varepsilon _{\mathrm{c}
}+\varepsilon _{\mathrm{p}}e^{-i\left( \omega _{\mathrm{p}}-\omega _{\mathrm{
c}}\right) t}=\kappa _{\mathrm{ex}}\left\langle \hat{c}\right\rangle,
\label{inputoutput}
\end{equation}
with the external loss rate $\kappa _{\mathrm{ex}}=\eta \kappa $. When the the coupling parameter  $\eta \ll 1$, the cavity is undercoupling, and when $\eta \simeq1$,
the cavity is overcoupled \cite{review4}.
Experimentally, $\eta $ can be continuously adjusted \cite{Vahala00,Vahala03}. 

Here, we concentrate on the component of the output field oscillating at the
probe frequency.  To study the phase-dependent optical response
properties for the probe field, we define the corresponding quadratures of the
field $\varepsilon _{\mathrm{T}}=\kappa _{\mathrm{ex}}\left\langle \delta 
\hat{c}\right\rangle /\varepsilon _{\mathrm{p}}$. The transmission
coefficient and power transmission coefficient can be further defined as $
\mathcal{T}=-1+\varepsilon _{\mathrm{T}}$ and $T=\left\vert \mathcal{T}\right\vert ^{2}$,
respectively. At weak cavity-waveguide coupling $\eta \ll 1$, $\left\vert
\mathcal{T}\right\vert \simeq 1-\mathrm{Re}\left( \varepsilon _{\mathrm{T}}\right) $, $
\mathrm{arg}\left(\mathcal{T}\right) \simeq -\mathrm{Im}\left( \varepsilon _{\mathrm{T
}}\right) $. Thus, similar to atomic physics, we can use the real and
imaginary parts of $\varepsilon _{\mathrm{T}}$ to represent absorptive and
dispersive behavior of the probe field. In the following, the ratio between $\left\vert \varepsilon _{\mathrm{a}}\right\vert$ and $\left\vert \varepsilon _{\mathrm{p}}\right\vert$ is defined as $y=\left\vert \varepsilon _{\mathrm{a}}/\varepsilon _{
\mathrm{p}}\right\vert$.

\section{\label{p3}Phase-dependent optical response properties for the probe field}

\subsection{\label{A}GWI-like absorption spectra}

Here we assume
$ \left\vert G\right\vert>\sqrt{\kappa \gamma _{\mathrm{m}}}/2$, i.e., the cooperativity $C=4\left\vert G\right\vert ^{2}/\left( \kappa
\gamma _{\mathrm{m}}\right)>1$. In this regime, one can abtain typical 
OMIT or Autler-Townes splitting spectra if only the control and the probe fields are applied. But if an additional
driving field is applied on the mechanical resonator, the interference between
 OMIT process and phonon-photon parametric process (represented by the first and the second terms in Eq.~\eqref{expectationC}, respectively) can lead to the expected phase-dependent absorption spectra.
In Figs.~\ref{EITA}(a)-\ref{EITA}(d), we plot absorption $\mathrm{Re}\left( \varepsilon _{\mathrm{T}}\right) $, dispersion 
$\mathrm{Im}\left( \varepsilon _{\mathrm{T}}\right) $, and power
transmission coefficient $T$ versus $\Delta^{\prime}$ for different relative phase $\Phi $. For simplicity, we have assumed the ratio of amplitude
between the two weak drivings $y=\left\vert \varepsilon _{\mathrm{a}}/\varepsilon _{
\mathrm{p}}\right\vert =1$. When $
\Phi =0$, the interference of the two terms in Eq.~\eqref{expectationC}
results in absorption and anomalous dispersion around $\Delta^{\prime }$= 0.
When $\Phi =\pi /2$, we can get asymmetric gain spectra with transparency
point at $\Delta ^{\prime }\simeq 0$ and absorption and amplification appear
in the red- and blue-detuned regions, respectively. The nature of
dispersion is normal in the transparency and amplification regions where
quantum interferences are prominent. When $\Phi =\pi $, a remarkable probe
gain can be established between two Autler-Townes absorption peaks, with the
maximum gain point being located at $\Delta ^{\prime }$= 0. The curve of $
\mathrm{Im}\left( \varepsilon _{\mathrm{T}}\right) $ exhibits normal
dispersive behavor in the amplification regime. When $\Phi =3\pi /2$, we
attain the mirror image of the $\Phi =\pi /2$ absorption curve. 

Note that Figs.~\ref
{EITA}(a)-\ref{EITA}(d) exhibit the similar type of phase-dependent GWI absorption spectra as those
in $\Delta $-type superconducting artificial atoms \cite{DeltaAA2,DeltaAA3}. But there also exist some differences between them. Specifically, a $\Delta $-type artificial atom is a three-level system, and one can easily check that when such an atom is driven by three coherent fields (i.e., a strong control, a weak probe, and an additional weak auxiliary field, respectively), 
the populations of the two levels related to the probe transition are
inversionless \cite{DeltaAA2,DeltaAA3}. While an optomechanical cavity is a system with infinite number of energy levels $\left\vert
N_{c},N_{m}\right\rangle$ [$N_{c}$($N_{m}$) denotes the number of photons (phonons)], the
probe field couples all the transitions $\left\vert
N_{c},N_{m}\right\rangle \leftrightarrow \left\vert
N_{c}+1,N_{m}\right\rangle $ [see Fig.~\ref{OM system}(b)], and the population-inversionless condition between these pairs of states is not necessarily satisfied. Thus we term the spectra in Fig.~\ref{EITA} as \emph{GWI-like absorption spectra}. 

\subsection{\label{B}Weak control field regime: OMIA and EIT-like spectra}

\begin{figure*}[t]
\includegraphics[width=1
\textwidth]{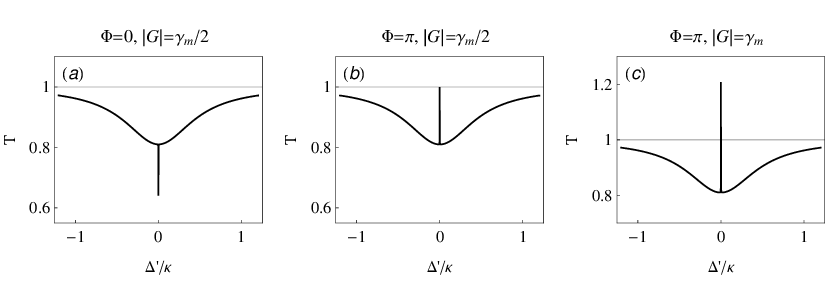}
\caption{(color online). Phase-dependent power transmission coefficient versus $\Delta ^{\prime }$ with $\Phi =0$, $\left\vert G\right\vert =
\protect\gamma _{\mathrm{m}}/2$ in (a); $\Phi =\protect\pi $, $
\left\vert G\right\vert =\protect\gamma _{\mathrm{m}}/2$ in (b); $
\Phi =\protect\pi $, $\left\vert G\right\vert =\protect\gamma _{\mathrm{m}}$
in (c). Other parameters are the same as in Fig.~\protect\ref
{EITA}.}
\label{EIA}
\end{figure*}

When $\left\vert G\right\vert \ll \sqrt{\kappa \gamma _{\mathrm{m}}}/2$,
i.e., the cooperativity $C\ll 1$, the expectation value of the
fluctuation operator $\delta \hat{c}$ can be approximately written as
\begin{equation}
\left\langle \delta \hat{c}\right\rangle =e^{i\phi _{\mathrm{p}}}\left[ 
\frac{\left\vert \varepsilon _{\mathrm{p}}\right\vert }{\frac{\kappa }{2}%
-i\Delta ^{\prime }}+\frac{2\left\vert \varepsilon _{\mathrm{a}}\right\vert
\left\vert G\right\vert e^{i\Phi }}{\kappa \left( \frac{\gamma _{\mathrm{m}}
}{2}-i\Delta ^{\prime }\right) }\right] .
\end{equation}
Clearly, the first term shows that in this parameter regime, the OMIT
effect vanishes, the probe absorption spectrum will exhibit usual
Lorentz line shape with width $\kappa $ in the absence of the driving field $
\varepsilon _{\mathrm{a}}$. However, in our case, due to the existence of $
\varepsilon _{\mathrm{a}}$, the photons generated by the phonon-photon
parametric process can interfere (depending
on the phase factor $\Phi$) with the photons directly exited by the probe beam.The absorptive behavior of the probe field can be represented by the real part of the quadrature of the field
\begin{equation}
\mathrm{Re}\left( \varepsilon _{\mathrm{T}}\right)
=\frac{\kappa_{\mathrm{ex}}}{2} \left[\frac{\kappa}{\frac{{\kappa}^2}{4}+{\Delta^\prime}^2}+\frac{2\left\vert G\right\vert\left(\gamma_{\mathrm{m}}\cos{\Phi}-2\Delta^\prime\sin{\Phi}\right)}{\kappa\left(\frac{\gamma_{\mathrm{m}}^2}{4}+{\Delta^\prime}^2\right)}\right],
\end{equation}
which depends on the phase factor $\Phi$.  Note that without loss of generality, we have let the ratio of amplitude between the two weak drivings equals to
one (i.e., $y=1$). Typically, when $\Phi=0$ or $\Phi=\pi$, we have
\begin{equation}
\mathrm{Re}\left( \varepsilon _{\mathrm{T}}\right)=\frac{\kappa_{\mathrm{ex}}}{2} \left[\frac{\kappa}{\frac{{\kappa}^2}{4}+{\Delta^\prime}^2}\pm\frac{2\left\vert G\right\vert\gamma_{\mathrm{m}}}{\kappa\left(\frac{\gamma_{\mathrm{m}}^2}{4}+{\Delta^\prime}^2\right)}\right].
\label{EIAandEIT}
\end{equation}
Here, the sign ``$+$" and ``$-$" correspond to relative phase  $\Phi=0$ and $\Phi=\pi$, respectively. Eq.~\eqref{EIAandEIT} is composed of a sum of two Lorentzians with width $\kappa$ and $\gamma _{\mathrm{m}}$, respectively. In addition, when $\Phi=0(\pi)$, constructive (destructive) interference appears. 
  In Figs.~\ref{EIA}(a)-\ref{EIA}(c) we plot power
transmission coefficient $T=1-2\mathrm{Re}\left( \varepsilon _{\mathrm{T}}\right)$ curves to display these kinds of  spectral structures resulting from phase-dependent constructive/destructive interference effects.

Specifically, when $
\Phi =0$, constructive interference occurs at $\Delta ^{\prime }=0$, 
resulting in typical OMIA spectrum with very sharp absorption feature around the resonant point, as shown in Figs.~\ref{EIA}(a). Note
that in optomechanical setups, similar OMIA spectrum for a
probe field can also be abtained by placing a pump blue-detuned at a mechanical
frequency away from cavity \cite{OMIT4,OMIA1}. Also, another version of OMIA was predicted in
a driving double-cavity configuration, where the absorption peak is
established in the OMIT window \cite{OMIA2}. When $\Phi =\pi $, destructive
interference occurs, thus a transparency or an amplification window can appear at the
resonance point, depending on the value of $\left\vert G\right\vert $.
According to Eq.~\eqref{EIAandEIT}, when $\left\vert G\right\vert =\gamma _{\mathrm{m}}/
2 $, the absorption $\mathrm{Re}\left( \varepsilon _{\mathrm{T}}\right)=0$ at resonant point. In this case, an EIT-like
power transmission curve can be abtained with the value of $T$ at the transparency dip being exactly
one, as shown in Figs.~\ref{EIA}(b). When $\left\vert G\right\vert >\gamma _{\mathrm{m}
}/2$, $\mathrm{Re}\left( \varepsilon _{\mathrm{T}}\right)<0$(i.e., $T>0$), a gain dip can be
established in the vicinity of cavity resonant point, as shown in Figs.~\ref{EIA}(c). 

Let us now make comparisons between the EIT-like phenomenon shown in Figs.~\ref{EIA}(b),
and the standard OMIT phenomenon \cite{OMIT1a,OMIT1b,OMIT1c,OMIT2,OMIT3,OMIT4,OMIT5}. In both cases, the coherent oscillation of the mechanical resonator induces
sidebands on the cavity field. Thus photons with frequency $\omega _{\mathrm{
p}}$ is generated and interfere destructively with the probe beam,
resulting in a sharp transparency window splitting the probe absorption peak. However, 
the coherent oscillation of the mechanical resonator is attributed to different mechanism in these two cases. In standard OMIT phenomenon, the mechanical resonator is driven by a time varying
radiation pressure force induced by the beat of the probe laser and the
control laser, and oscillates coherently. To manifest this effect, a relatively large effective optomechanical coupling constant with $
\left\vert G\right\vert \gtrsim \sqrt{\kappa \gamma _{\mathrm{m}}}/2$ (i.e., 
$C\gtrsim 1$) is required. While in present EIT-like case, $\left\vert G\right\vert
=\gamma _{\mathrm{m}}/2 \ll \sqrt{\kappa \gamma _{\mathrm{m}}}
/2$, the usual OMIT effect already vanishes, but the mechanical resonator is still directly driven 
by the external driving field with amplitude $\varepsilon _{\mathrm{a}}$ and oscillates
coherently. Thus the EIT-like effect may provide an alternative way to control photon propagation even if the control field
is too weak to produce usual OMIT phenomenon. Note that in a recent experiment on coherent signal transfer between microwave and optical fields, this type of phenomenon has
been used to demonstrate coherent interactions between microwave, mechanical
and optical modes \cite{OMdr1}. 

\begin{figure*}[t]
\includegraphics[width=1
\textwidth]{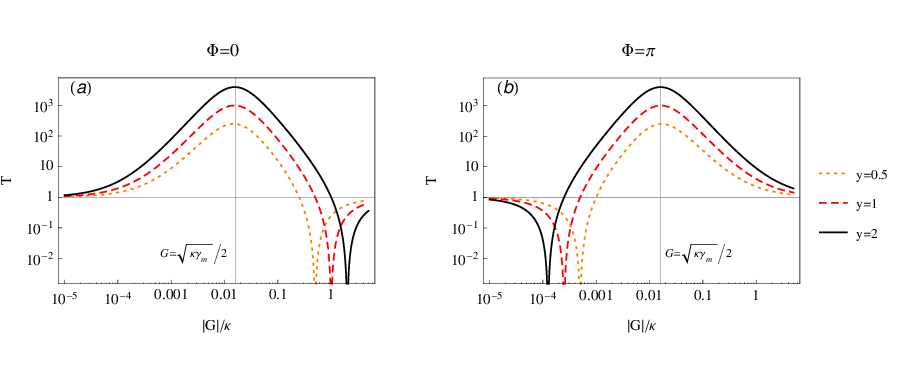}
\caption{(color online). The power transmission coefficient
for the probe as function of $\left\vert G\right\vert $ for
various $y$, is plotted with $\Phi=0$ in (a) and $\Phi=\pi$ in (b). In all cases, the cavity-waveguide coupling parameter
is $\protect\eta =1$. Other parameters are the same as in Fig.~\protect
\ref{EITA}.}
\label{amplifier}
\end{figure*}

\subsection{\label{C}Amplification and perfect absorption for the probe beam}

Usually, an amplifier based on optomechanical
setup is realized by pumping the optomechnical cavity by a blue-detuned control field \cite{OMIT4,OMIT5,OMAmplification1}. Our proposal
shows that a red-detuned control field associating with an auxiliary driving applied to the mechanical resonator can also realize
probe amplification. In previous subsections, to get power transmission spectra analogous to those  investigated in atomic gases (such as EIA, GWI), we have taken
coupling parameter $\eta \ll 1$, and have shown a gain dip around $\Delta ^{\prime }=0$ when $\Phi=\pi$ [see
Fig.~\ref{EITA}(c) and Figs.~\ref{EIA}(c)]. Here, to abtain a remarkable amplification
for a resonantly injected probe, we take $\eta =1$ (i.e., the cavity is
over-coupled) and $\Phi $ is equal to either $0$ or $\pi $. Substituting Eq.~\eqref{expectationC} into relation $T=\left\vert -1+\epsilon_{\mathrm{T}}\right\vert^2$ and letting $\Delta^\prime=0$, we get the power transmission coefficient at resonant point as function of $\left\vert G\right\vert$: 
\begin{equation}
T=\left(\frac{\frac{\kappa\gamma_{\mathrm{m}}}{4}\pm y\kappa\left\vert G\right\vert-\left\vert G\right\vert^2}{\frac{\kappa\gamma_{\mathrm{m}}}{4}+\left\vert G\right\vert^2}\right)^2.
\label{amp}
\end{equation}
Here, the sign ``$+$" and ``$-$" correspond to relative phase  $\Phi=0$ and $\Phi=\pi$, respectively. From Eq.~\eqref{amp},  we can find that a resonant probe can be effectively amplified, the main results can be  summarized as follows: (i)
when $\Phi =0$, under the contition $\kappa\ll\gamma_{\mathrm{m}}$, as is often the case in cavity optomechanics, the
amplification region (with $T>1$) is approximately $\left\vert G\right\vert <y\kappa /2$;
(ii) when $\Phi =\pi $, the amplification region is approximately $\left\vert G\right\vert >\gamma _{
\mathrm{m}}/\left( 2y\right) $; (iii) in both $\Phi =0$ and $\Phi =\pi $ cases, 
when $\left\vert G\right\vert \simeq \sqrt{\kappa \gamma _{\mathrm{m}}}/2$
(i.e., the cooperativity $C\simeq 1$), for different ratio $y$ between the
amplitudes of the two weak fields, $\mathrm{d} T/\mathrm{d}\vert G\vert=0$, $\mathrm{d}^2 T/\mathrm{d}\vert G\vert^2<0$, the output power for the field at
probe frequency $\omega _{\mathrm{p}}$ achieves maximum with power
transmission coefficient $T_{\mathrm{max}}\simeq y^{2}\kappa /\gamma _{\mathrm{
m}}$. Figs.~\ref{amplifier}(a) and \ref{amplifier}(b) show the amplitude of the output power from the cavity as
function of $\left\vert G\right\vert $ in these cases.

Physically, the
extra energy of the amplified probe is due to the contribution of the
phonon-photon parametric process described by the second term in Eq.~\eqref{expectationC},
whose strength is dependent on the coherent photons (excited by $\varepsilon
_{\mathrm{c}}$) in the cavity and phonons (excited by $\varepsilon _{\mathrm{a}}$) in the mechanical resonator. On one hand, for a given probe, increasing 
$y$ (by increasing $\left\vert \varepsilon _{\mathrm{a}}\right\vert $) can
excite more phonons in the mechanical resonator, leading to a more remarkable amplification,
as shown in Figs.~\ref{amplifier}(a) and \ref{amplifier}(b). On the other hand, an
increasing $\left\vert G\right\vert $ (by increasing $\left\vert \varepsilon _{\mathrm{c}
}\right\vert $) can produce more photons in the cavity but at the same time
lower the phonon numbers in the mechanical resonator for the existence of
sideband cooling effect. The first process contributes positively and the
second one negatively to the phonon-photon parametric process, resulting in
maximal amplification appearing at $\left\vert G\right\vert \simeq \sqrt{\kappa
\gamma _{\mathrm{m}}}/2$, as shown in Fig.~\ref{amplifier}(a) and \ref{amplifier}(b). Note that at the maximal amplification
point, the modulus of the expectation value $\left\vert \left\langle \delta \hat{c}
\right\rangle \right\vert$ may be very large, to ensure
the validity of the linearize theory, $\left\vert \left\langle \delta \hat{c}
\right\rangle \right\vert /c_{\mathrm{s}}\ll 1$ should be satisfied. Using
this relation, we can estimate that the condition $\sqrt{T_{\mathrm{max}}}
\left\vert \varepsilon _{\mathrm{p}}/\varepsilon _{\mathrm{c}}\right\vert
\ll 1$ (i.e., the probe magnitude after amplification must have a lower
value than that of the control field) must be satisfied to ensure the
validity of the linear-regime analysis.

In addition, Eq.~\eqref{amp} shows that when 
$\Phi =0$ and $\left\vert G\right\vert
\simeq y\kappa$, or $\Phi =\pi$ and $\left\vert G\right\vert \simeq \gamma _{
\mathrm{m}}/\left( 4y\right)$, the power transmission coefficient for a resonant injected probe
beam is zero. This means that the probe can be totally absorbed. These results can be clearly seen in Figs.~\ref{amplifier}(a) and \ref{amplifier}(b). It is
known that for over-coupled case $\eta =1$, if a single probe laser drives the
cavity, the output probe beam will leave
the cavity without any absorption \cite{review3}. However, in our case, because
a control field $\varepsilon _{\mathrm{c}}$ and an auxiliary driving field $\varepsilon _{\mathrm{a}}$ are applied, destructive interference can lead
to zero output for the probe field. Thus, the device may be used as a
quantum switch to control the photon propagation in the future quantum
network. We note that similar perfect absorption
phenomena also exist in two-side driving resonator-in-middle type optomechanical systems
\cite{PerfectAbsorption1,PerfectAbsorption2}.
\begin{figure*}[t]
\includegraphics[width=0.8
\textwidth]{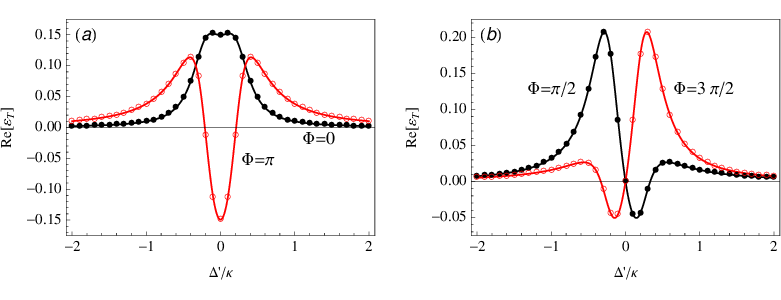}
\caption{(color online). Comparison between the numerical (dots and circles) and the
analytical (solid curves)\ results of the phase-dependent absorption spectra
with (a) $\Phi =0,\protect\pi $; (b) $\Phi =\protect\pi /2,3\protect\pi /2$.
The average thermal phonon number $N_{\mathrm{th}}=10$. The coupling strength 
$\left\vert G\right\vert =\protect\kappa /3$. The probe amplitude $\left\vert \protect\varepsilon _{
\mathrm{p}}\right\vert =\protect\kappa /30$. Other parameters are the same
as in Fig.~\protect\ref{EITA}.}
\label{numerical}
\end{figure*}

\subsection{\label{D}Numerical results}

In this part, to verify the above phase-dependent spectral structure
abtained analytically, we give numerical results by solving the master equation. The quantum Langevin equations 
\eqref{linearLangevinRWA1} and \eqref{linearLangevinRWA2} correspond to an
effective Hamiltonian 
\begin{eqnarray}
\hat{H}_{\mathrm{eff}} &=&-\hbar \Delta ^{\prime }\left( \delta \hat{c}%
^{\dagger }\delta \hat{c}+\delta \hat{b}^{\dagger }\delta \hat{b}\right)
-\left( \hbar G\delta \hat{c}^{\dagger }\delta \hat{b}+\hbar G^{\ast }\delta 
\hat{c}\delta \hat{b}^{\dagger }\right)   \notag \\
&&+\left( i\hbar \varepsilon _{\mathrm{p}}\delta \hat{c}^{\dagger }+i\hbar
\varepsilon _{\mathrm{a}}\delta \hat{b}^{\dagger }+\mathrm{H.c.}\right),
\end{eqnarray}
with beam-splitter like interaction. Based on this Hamiltonian, we can get
the quantum master equation
\begin{eqnarray}
\dot{\hat{\rho}} &=&\frac{1}{i\hbar }\left[ \hat{H}_{\mathrm{eff}},\hat{\rho}%
\right] +\kappa \mathcal{D}\left[ \delta \hat{c}\right] \hat{\rho}+\gamma _{%
\mathrm{m}}\left( N_{\mathrm{th}}+1\right) \mathcal{D}\left[ \delta \hat{b}%
\right] \hat{\rho}  \notag \\
&&+\gamma _{\mathrm{m}}N_{\mathrm{th}}\mathcal{D}\left[ \delta \hat{b}%
^{\dagger }\right] \hat{\rho}
\end{eqnarray}
describing the dynamics of system, where $\hat{\rho}$ denotes the density
matrix of the system, $\mathcal{D}\left[ \hat{o}\right] \hat{\rho}=\hat{o}
\hat{\rho}\hat{o}^{\dagger }-\left( \hat{o}^{\dagger }\hat{o}\hat{\rho}+\hat{
\rho}\hat{o}^{\dagger }\hat{o}\right) /2$ ($\hat{o}=\delta \hat{c},\delta 
\hat{b},\delta\hat{b}^{\dagger }$) is the standard dissipator in Lindblad form, and $N_{\mathrm{th}}$
is the average thermal phonon number of the mechanical resonator. For a
nanomechanical resonator with frequency $2\pi\times$10MHz, under typical environment
temperature (30mK) in present experiments \cite{OMAmplification1}, the thermal
phonon number $N_{\mathrm{th}}$ is about 10. Fig.~\ref{numerical}
gives both the numerical and analytical results of the phase-dependent probe
absorption spectra. Without loss of generality, we only take GWI-like case discussed in subsection \ref{A} as an example. 
We can see that the analytical results is in good
agreement with the numerical calculations.

\section{\label{p4}Conclusions and discussions}

In summary, we have explored an optomechanical system under EIT condition with the mechanical resonator being driven by an auxiliary coherent field. We find that the response of the driven optomechanical system to the weak probe field depends on the total phase of three classical fields.
Because an additional driving field is applied to the mechanical resonator, the system will
exhibit more complex
quantum interference phenomena. Specifically, when the cooperativity $C=4\left\vert G\right\vert ^{2}/\left( \kappa
\gamma _{\mathrm{m}}\right)>1$ is satisfied, we can get GWI-like spectra similar to those predicted in superconducting artificial
atoms. When the cooperativity  $C\ll1$, our proposal provide a way to abtain OMIA and EIT-like spectra. When
the coperativity $C\simeq1$, we can get remarkable amplification for the probe beam by adjusting the phase and amplitude of the coherent driving field applied on the mechanical resonator. We also give numerical results
including thermal decoherence by solving the master equation. The numerical
results are in good agreement with the analytical ones. Experimentally,
there are various ways to coherently drive a micro/nano scale mechanical
resonator \cite{OMdr1,OMdr2,OMdr3}. This kind of optomechanical setups may be used to switch or amplify probe signals in the future quantum networks.

\begin{acknowledgments}
We thank Xiao-Bo Yan and Wei Nie for useful discussions. W. Z. J. is supported by the NSFC under Grant No.~11347001,
￼￼
￼￼11404269, and the Fundamental Research Funds for the Central Universities 2682014RC21. L. F. W. is supported by the NSFC under Grant
Nos.~11174373, U1330201, and 91321104. Y. X. L. is supported by the NSFC under Grant
Nos.~61025022 and 61328502, the National Basic Research Program of China 973 Program
under Grant No. 2014CB921401, the Tsinghua
University Initiative Scientific Research Program, and
the Tsinghua National Laboratory for Information Science
and Technology (TNList) Cross-discipline Foundation. It is also supported by NSFC under Grant No.~11174027.
\end{acknowledgments}



\begin{thebibliography}{9}

\bibitem{review1} T. J. Kippenberg and K. J. Vahala, Science \textbf{321}, 1172 (2008).
\bibitem{review2} F. Marquardt and S. M. Girvin, Physics \textbf{2}, 40 (2009).
\bibitem{review3} M. Aspelmeyer, P. Meystre, and K. Schwab, Phys. Today \textbf{65}(7), 29 (2012).
\bibitem{review4} M. Aspelmeyer, T. J. Kippenberg, and F. Marquardt, Rev. Mod. Phys. \textbf{86}, 1391 (2014).
\bibitem{EIT1} S. E. Harris, J. E. Field, and A. Imamoglu, Phys. Rev. Lett. \textbf{64}, 1107 (1990).
\bibitem{EIT2} S. E. Harris, Phys. Today \textbf{50}, 36 (1997).
\bibitem{EIT3} M. Fleischhauer, A. Imamoglu, and J. P. Marangos, Rev. Mod. Phys. \textbf{77}, 633 (2005).
\bibitem{OMIT1a} G. S. Agarwal and S. Huang, Phys. Rev. A \textbf{81}, 041803 (2010).
\bibitem{OMIT1b} S. Huang and G. S. Agarwal, Phys. Rev. A \textbf{83}, 023823 (2011).
\bibitem{OMIT1c} S. Huang and G. S. Agarwal, Phys. Rev. A \textbf{83}, 043826 (2011).
\bibitem{OMIT2} S. Weis, R. Rivi\`{e}re, S. Del\'{e}́glise, E. Gavartin, O. Arcizet, A. Schliesser, and T.J. Kippenberg, Science \textbf{330}, 1520 (2010).
\bibitem{OMIT3} J. D. Teufel, D. Li, M. S. Allman, K. Cicak, A. J. Sirois,J. D. Whittaker, and R. W. Simmonds, Nature (London), \textbf{471}, 204 (2011).
\bibitem{OMIT4} A. H. Safavi-Naeini, T. P. M. Alegre, J. Chan, M. Eichenfield, M. Winger, Q. Lin, J. T. Hill, D. E. Chang, and O. Painter, Nature (London) \textbf{472}, 69 (2011).
\bibitem{OMIT5} M. Karuza, C. Biancofiore, M. Bawaj, C. Molinelli, M. Galassi, R. Natali, P. Tombesi, G. Di Giuseppe, and D. Vitali, Phys. Rev. A \textbf{88}, 013804 (2013).
\bibitem{OMslowlight1} X. Zhou, F. Hocke, A. Schliesser, A. Marx, H. Huebl, R. Gross, and T. J. Kippenberg, Nat. Phys. \textbf{9}, 179, (2013). 
\bibitem{OMslowlight2} D. E. Chang, A. H. Safavi-Naeini, M. Hafezi, and O. Painter, New J. Phys. \textbf{13}, 023003 (2011).
\bibitem{nolinearOMIT1} Hao Xiong, Liu-Gang Si, An-Shou Zheng, Xiaoxue Yang, and Ying Wu, Phys. Rev. A \textbf{86}, 013815 (2012).
\bibitem{nolinearOMIT2} A. Kronwald and F. Marquardt, Phys. Rev. Lett. \textbf{111}, 133601 (2013).
\bibitem{nolinearOMIT3} K. B{\o}rkje, A. Nunnenkamp, J. D. Teufel, and S. M. Girvin, Phys. Rev. Lett. \textbf{111}, 053603 (2013).
\bibitem{nolinearOMIT4} M.-A. Lemonde, N. Didier, and A. A. Clerk, Phys. Rev. Lett. \textbf{111}, 053602 (2013).
\bibitem{EIA1}  A. Lezama, S. Barreiro, and A. M. Akulshin, Phys. Rev. A \textbf{59}, 4732 (1999); A. M. Akulshin, S. Barreiro, and A. Lezama, Phys. Rev. A \textbf{57}, 2996 (1998); A. Lipsich, S. Barreiro, A. M. Akulshin, and
A. Lezama, Phys. Rev. A \textbf{61}, 053803 (2000).
\bibitem{EIA2} A. V. Taichenachev, A. M. Tumaikin, and V. I. Yudin, Phys. Rev. A \textbf{61}, 011802 (1999). 
\bibitem{EIA3} Hou Ian, Yu-xi Liu, and F. Nori, Phys. Rev. A \textbf{81}, 063823 (2010).
\bibitem{OMIA1} F. Hocke, X. Zhou, A. Schliesser, T. J. Kippenberg, H. Huebl, and R. Gross, New J. Phys. \textbf{14}, 123037 (2012).
\bibitem{OMIA2} Kenan Qu and G. S. Agarwal, Phys. Rev. A \textbf{87},  031802 (2013).
\bibitem{OMAmplification1} F. Massel, T. T. Heikkil\"{a}̈, J.-M. Pirkkalainen, S. U. Cho, H. Saloniemi, P. J. Hakonen, and M. A. Sillanp\"{a}\"{a}̈, Nature (London) \textbf{480}, 351 (2011).
\bibitem{OMAmplification2} A. Metelmann and A. A. Clerk, Phys. Rev. Lett. \textbf{112}, 133904 (2014).
\bibitem{OMAmplification3} A. Nunnenkamp, V. Sudhir, A. K. Feofanov, A. Roulet, and T. J. Kippenberg, Phys. Rev. Lett. \textbf{113}, 023604 (2014).
\bibitem{AtomloopScully1} M. O. Scully, S. Y. Zhu, and A. Gavrielides, Phys. Rev. Lett. \textbf{62}, 2813
(1989).
\bibitem{AtomloopScully2} H. Fearn, C. Keitel, M. O. Scully and S. Y. Zhu, Opt. Commun. \textbf{87}, 323 (1992).
\bibitem{AtomloopScully3} H. Li, V. A. Sautenkov, Y. V. Rostovtsev, G. R. Welch, P. R. Hemmer, and M. O. Scully, Phys. Rev. A \textbf{80}, 023820 (2009).
\bibitem{SGC} S. Menon and G. S. Agarwal, Phys. Rev. A \textbf{57}, 4014 (1998).
\bibitem{phaseonium1} M.O. Scully, Phys. Rep. \textbf{219}, 191 (1992).
\bibitem{phaseonium2} 
M. Fleischhauer, C. H. Keitel, M. O. Scully, C. Su, B. T. Ulrich, and S. Y. Zhu, Phys. Rev. A \textbf{46}, 1468 (1992).
\bibitem{CorrelatedLasing1} 
M. O. Scully, K. W\'{o}dkiewicz, M. S. Zubairy, J. Bergou, N. Lu, and J. Meyer ter Vehn, Phys. Rev. Lett. \textbf{60}, 1832(1988).
\bibitem{CorrelatedLasing2} 
M. O. Scully and M. S. Zubairy, \textit{Quantum
optics}, (Cambridge University Press, Cambridge, 1997).
\bibitem{AtomloopKnight} J. C. Petch, C. H. Keitel, P. L. Knight, and J. P. Marangos, Phys. Rev. A \textbf{53}, 543 (1996).
\bibitem{AtomloopKosachiov} D. V. Kosachiov, B. G. Matisov, and Yu. V. Rozhdestvensky, Opt. Commun. \textbf{85}, 209 (1991); J. Phys. B \textbf{25}, 2473 (1992).
\bibitem{OMdr1} A. D. O'Connell, M. Hofheinz, M. Ansmann, R. C. Bialczak, M. Lenander, E. Lucero, M. Neeley, D. Sank, H. Wang, M. Weides, J. Wenner, J. M. Martinis, and A. N. Cleland, Nature (London) \textbf{464}, 697 (2010).
\bibitem{OMdr2} J. Bochmann, A. Vainsencher, D. D. Awschalom and A. N. Cleland, Nat. Phys. \textbf{9}, 712 (2013).
\bibitem{OMdr3} Yu-xi Liu, A. Miranowicz, Y. B. Gao, J. Bajer, C. P. Sun, and F. Nori, Phys. Rev. A \textbf{82}, 032101 (2010).
\bibitem{DeltaAA1} Yu-xi Liu, J. Q. You, L. F. Wei, C. P. Sun, and F. Nori, Phys. Rev. Lett. \textbf{95}, 087001 (2005).
\bibitem{DeltaAA2} W. Z. Jia and L. F. Wei, Phys. Rev. A \textbf{82}, 013808 (2010).
\bibitem{DeltaAA3}  J. Joo, J. Bourassa, A. Blais, and B. C. Sanders. Phys. Rev. Lett. \textbf{105}, 073601
(2010).
\bibitem{DeltaAA4} W. Z. Jia, L. F. Wei, and Z. D. Wang, Phys. Rev. A \textbf{83}, 023811 (2011).
\bibitem{DeltaMolecule1} P. Kra’l and M. Shapiro, Phys. Rev. Lett. \textbf{87}, 183002 (2001); P. Kra’l, I. Thanopulos, M. Shapiro, and D. Cohen, Phys. Rev. Lett. \textbf{90}, 033001 (2003).  
\bibitem{DeltaMolecule2} Y. Li, C. Bruder, and C. P. Sun, Phys. Rev. Lett. \textbf{99}, 130403 (2007).
\bibitem{inputoutput} D. F. Walls and G. J. Milburn, \textit{Quantum Optics} (Springer-Verlag, Berlin, 1994).
\bibitem{Vahala00} M. Cai, O. Painter, K. J. Vahala, Phys. Rev. Lett. \textbf{85}, 74 (2000).
\bibitem{Vahala03} S. M. Spillane, T. J. Kippenberg, O. J. Painter, K. J. Vahala, Phys. Rev. Lett. \textbf{91}, 043902 (2003).
\bibitem{PerfectAbsorption1} G. S. Agarwal and S. Huang, New J. Phys. \textbf{16}, 033023 (2014).
\bibitem{PerfectAbsorption2} X. B. Yan, C. L. Cui, K. H. Gu, X. D. Tian, C. B. Fu, and J. H. Wu, Opt. Express \textbf{22}, 4886 (2014).

\end{thebibliography}
\end{document}